\documentclass[10pt,letter]{article}
\usepackage{multicol}
\usepackage{multirow}
\usepackage{hhline}
\usepackage{graphicx}
\usepackage[T1]{fontenc}
\usepackage{pslatex}
\usepackage{subfigure}
\usepackage{fancyvrb}
\usepackage{tabularx}
\usepackage{multirow}
\usepackage{booktabs}
\usepackage{dcolumn}
\usepackage{enumerate}
\usepackage{verbatim}
\usepackage{mathtools}
\usepackage{url}
\usepackage{rotating}

\usepackage[margin=0.8in]{geometry}
\usepackage{xcolor}

\usepackage{etoolbox}
\makeatletter
\patchcmd{\maketitle}{\@copyrightspace}{}{}{}
\makeatother

\begin{document}

\title{Co-evolutionary dynamics in social networks:\\A case study of Twitter}

\date{}

\author{
Demetris Antoniades\\
       College of Computing\\
       Georgia Institute of Technology\\
       danton@gatech.edu
\and
Constantine Dovrolis\\
       College of Computing\\
       Georgia Institute of Technology\\
       constantine@gatech.edu
}

\maketitle
\begin{abstract}
Complex networks often exhibit co-evolutionary dynamics,
meaning that the network topology and the state of nodes or links are coupled,
affecting each other in overlapping time scales. 
We focus on the co-evolutionary dynamics of
online social networks, and on Twitter in particular.
Monitoring the activity of thousands of Twitter users in real-time, 
and tracking their followers and tweets/retweets, we propose a method
to infer new retweet-driven follower relations.
The formation of such relations is much more likely
than the exogenous creation of new followers in the absence of any retweets.
We identify the most significant factors 
(reciprocity and the number of retweets
that a potential new follower receives) and propose a
simple probabilistic model of this effect.
We also discuss the implications of such co-evolutionary
dynamics on the topology and function of a social network.
\end{abstract}

Online Social Networks (OSNs), such as Twitter and Facebook, have changed
how individuals interact with society, how information flows between
actors, and how people influence each other.   
These are all complex dynamic processes that are now widely studied
empirically and in large scale, thanks to the availability of data from OSNs. 
Most OSN studies 
focus on one of the following two aspects of network dynamics. 
Dynamics {\em on} networks refer to changes in the state of network
nodes or links considering a static 
topology~\cite{bakshy2009social,vespignani2011modelling}. 
Dynamics {\em of} networks, on the other hand,
refer to changes in the topology of a network,
without explicitly modeling its underlying 
causes~\cite{Leskovec:2005:GOT:1081870.1081893}. 
As noted by Gross and Blasius in~\cite{gross2008adaptive}, however,
real OSNs typically exhibit both types of dynamics, 
forming an adaptive, or co-evolutionary, system in which the network topology
and the state of nodes/links affect each other through a (rather poorly 
understood) feedback loop.

Dynamic processes in OSNs, such as information diffusion or influence, 
are obviously affected by the underlying network topology, but they 
also have the power to affect that topology. 
For instance, users may decide to add or drop a ``friendship'' or ``follower''
relation depending on what the potential ``friend'' or ``followee'' has
recently said or done in the context of that OSN. 
Previous empirical or modeling OSN studies often choose to ignore
such co-evolutionary dynamics, mostly for simplicity, 
assuming a static network topology, or assuming that the
topology and node/link states are decoupled and evolve 
in separate time scales \cite{DBLP:conf/sdm/LeskovecMFGH07}.

\begin{figure}[t]
\centering
	\includegraphics[width=\columnwidth,height=0.2\columnwidth]{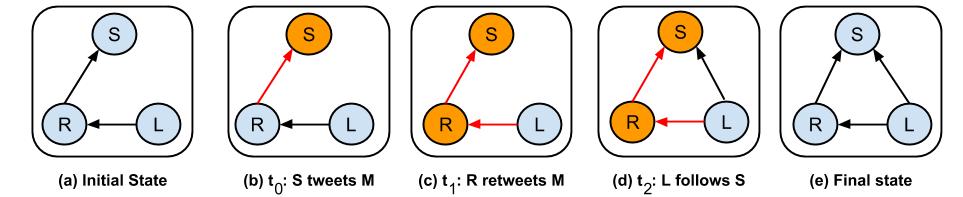}
	\caption{Network co-evolution: a Tweet-Retweet-Follow event.}
	\label{fig:TRFsteps}
\end{figure}

The literature on co-evolutionary dynamics has relied mostly
on abstract models so far, without sufficient empirical validation. 
For instance, Kosma and Barrat examined how the topology 
of an adaptive network of interacting agents and of the agents' opinions 
can influence each other~\cite{kozma2008consensus}. 
When agents rewire their links in a way that depends on the opinions of 
their neighbors, the result can be either a large number of small clusters, 
making global consensus difficult, 
or a highly connected but polarized network. 
Shaw and Schwartz~\cite{PhysRevE.81.046120} 
examined the effects of vaccination in static versus adaptive networks. 
Interestingly, they show that vaccination is much more 
effective in adaptive networks, and that two orders of magnitude 
less vaccine resources are needed in adaptive networks.
Volz and Mayers studied epidemics in dynamic contact 
networks and showed that the rate at which contacts are 
initiated and terminated affects the disease reproductive 
ratio~\cite{volz2009epidemic}. They concluded that 
static approximations of dynamic networks can be inadequate.
Rocha et al. simulated epidemics in an empirical 
spatio-temporal network of sexual contacts~\cite{rocha2011simulated},
showing that dynamic network effects accelerate epidemic outbreaks.
Perra et al. studied the effect of time-varying networks in random walks 
and search processes~\cite{PhysRevLett.109.238701}. The behavior 
of both processes was found to be ``strikingly different'' compared to 
their behavior in static networks.

The most relevant prior work, by Weng et al., 
analyzed the complete graph and activity 
of \textit{Yahoo! Meme}, a currently inactive Twitter-like social network,
to identify the effect of information diffusion on the evolution of the
underlying network~\cite{Weng:2013:RID:2487575.2487607}. 
They show that information diffusion causes about 24\% of 
the new links, and that the likelihood of a new link created by 
a user X to a user Y increases with the number of Y's messages seen by X. 

In this paper, we focus on co-evolutionary dynamics in the context of Twitter.
Twitter users create {\em follower-followee} relations with each other. 
A directed link from a user $R$ to a user $S$,
denoted by $R \rightarrow S$, means that $R$
is a follower of $S$, receiving $S$'s tweets; $S$ is referred to
as a followee of $R$.
$R$ can choose to propagate a tweet of $S$ to her own followers,
denoted by $F(R)$, creating a {\em retweet}.
When a follower $L \in F(R)$ receives a retweet of $S$ through $R$,
$L$ can choose to add $S$ to her followers. 
We call this sequence a {\em Tweet-Retweet-Follow (TRF)} event,
and refer to its three main actors as {\em Speaker} $S$,
{\em Repeater} $R$, and {\em Listener} $L$.
TRF events represent a clear case of co-evolutionary dynamics: 
information propagation (tweet-retweet) causes a 
topology change (new follower).

Figure~\ref{fig:TRFsteps} shows this sequence of events for the
simplest TRF case in which $R \rightarrow S$ and $L \rightarrow R$. 
In general, the Repeater $R$ may not be a follower of $S$ but 
she may receive $S$'s tweet through a cascade of retweets.
Additionally, the Listener $L$ may receive multiple retweets
of $S$ from the same or from different Repeaters.

The contributions of this study are:
\begin{enumerate}
\item We propose a measurement approach to detect TRF events, based
on near real-time monitoring of a Speaker's activity and followers.
\item We show that the formation of new follower relations through
TRF events is orders of magnitude more likely than the exogenous
arrival of new followers in the absence of any retweets. 
\item We identify the most significant factors for the 
likelihood of a TRF event: reciprocity (i.e., is Speaker $S$ 
already following Listener $L$?), number of received retweets
(i.e., how many retweets of $S$ were received at $L$ during
a given time interval $\Delta$), and of course the interval $\Delta$
itself.
\item We propose a simple but accurate two-parameter 
model to capture the probability of TRF events.
\item We discuss the implications of TRF events in the
structure and function of social networks.
\end{enumerate}

\section*{Results}

\subsection*{Endogenous versus exogenous link creation}
\label{sec:trfexo}

A user also gains new followers due to exogenous factors, 
such as Twitter's ``Who to follow'' service~\cite{guptawtf}.
Here, we compare the likelihood with which a user gains new followers 
when there are no recent retweets of her messages 
(exogenous link creation) compared to the case that she gains
new followers when at least one of her messages
has been recently retweeted (endogenous link creation). 

\begin{figure}[t]
	\centering
	\subfigure[] {
		\includegraphics[width=0.45\columnwidth,height=0.2\columnwidth]{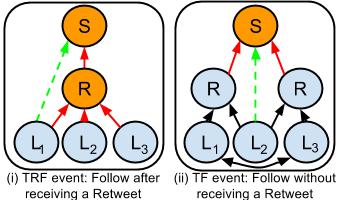}
		\label{fig:trfVStwf}
	}
	\subfigure[] {
		\includegraphics[width=0.45\columnwidth,height=0.2\columnwidth]{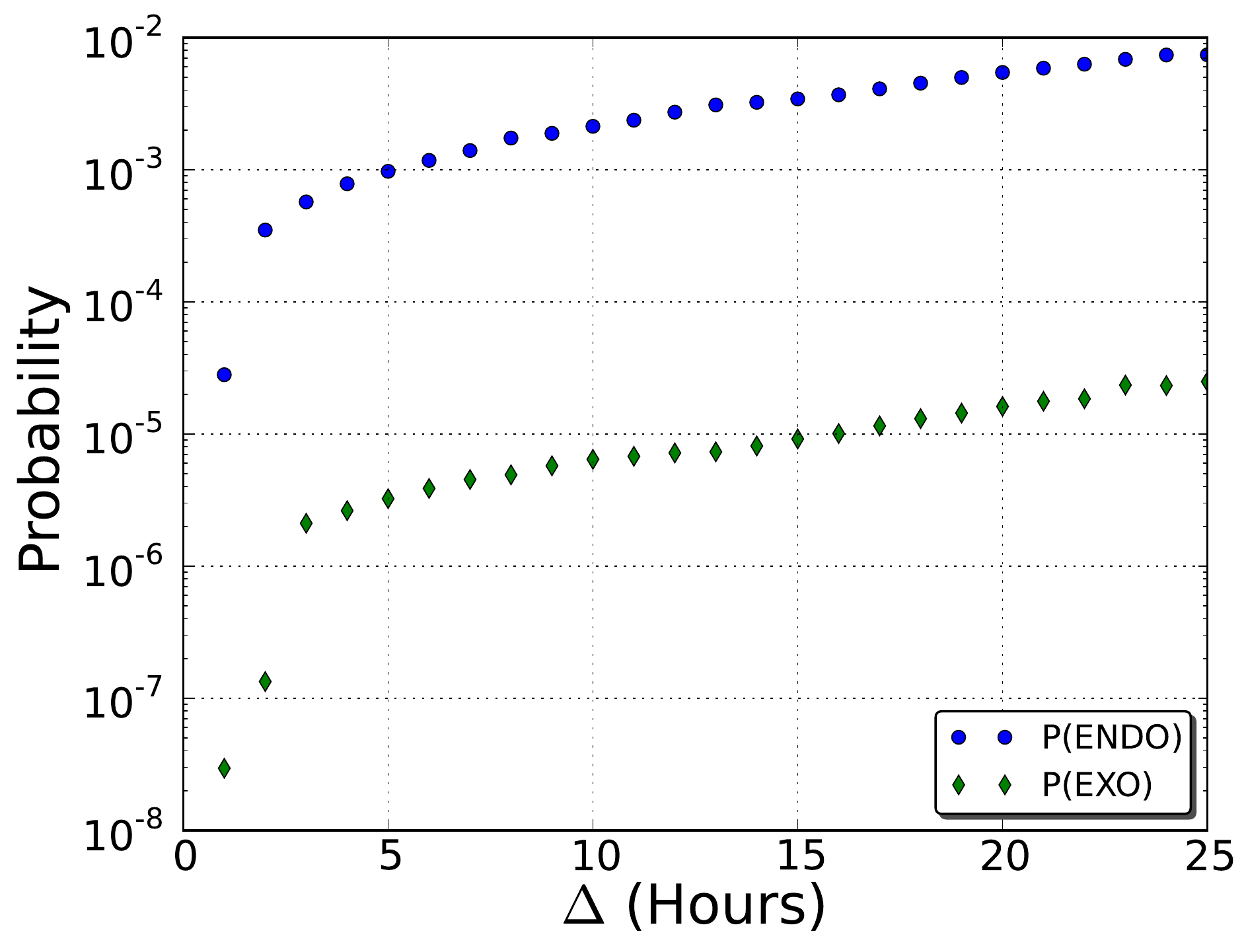}
		\label{fig:ptrfVSptwf}
	}
	\caption{~\subref{fig:trfVStwf} Controlling for the structural relation between $S$, $R$ and $L$ and for the activity of $S$ allows us to compare the likelihood of a new follower $L$ when $L$ received a retweet of $S$ $(i)$ compared to the case that $L$ did not receive a retweet of $S$ $(ii)$. The arrow direction shows who follows whom. Orange nodes represent tweet or retweet activity. Red edges show the extent of information propagation. Green dashed edges show new follower links.
		~\subref{fig:ptrfVSptwf} Probability that a Speaker $S$ gains at least one new 
	follower $L$ within an interval $\Delta$ from the time of a tweet (TF) or retweet (TRF) 
	of $S$. The Listener $L$ is not a follower of $S$ at the time of the 
	tweet (TF) or retweet (TRF).}
	\label{}
\end{figure}

We focus here on potential new followers $L$ of $S$ that were already following 
a follower of $S$. That is, we only examine three-actor relations 
in which $L \rightarrow R$ and  $R \rightarrow S$.
We then ask 
{\em ``is it more likely that $L$ will follow $S$ ($L \rightarrow S$)
when $L$ received a retweet of $S$ through $R$ (TRF event) or 
when $L$ did not receive any retweet from her followees
that follow $S$ (TF}
{\em  event)?''}
Figure~\ref{fig:trfVStwf} illustrates the TRF and TF events.
Note that the difference between endogenous (TRF) and exogenous (TF)
events is the retweet of $S$ from $R$; 
the local structure and the activity of $S$ remain the same in both cases.

We estimate the probability $P_{EXO}(\Delta)$ 
of exogenous new followers as follows.
Consider a tweet of Speaker $S$ at time $t_s$.
Suppose that this tweet is not retweeted by any of the followers of $S$
in the period $[t_s, t_s+\Delta]$.
Let $\Phi(S,t_s)$ be the set of followers of followers of $S$
that are not directly following $S$ at $t_s$, i.e.,
$\Phi(S,t_s)=\{X: X\not\in F(S,t_s), X\in F(Y,t_s), Y\in F(S,t_S)\}$. 
What is the fraction of these users that 
follow $S$ by time $t_s+\Delta$?

\begin{equation}
	P_{EXO}(\Delta) = \frac{|L: L\in \Phi(S, t_s), L\in F(S, t_s + \Delta)|}{|\Phi(S,t_s)|}
\end{equation}

Similarly, we estimate the probability $P_{ENDO}(\Delta)$ of 
endogenous new followers as follows.
Consider again a tweet of Speaker $S$ at time $t_s$ but suppose that 
this message has been retweeted by a specific follower of $S$,
referred to as Repeater $R$, at time $t_r>t_s$. 
Let $\Phi_R(S,t_r)$ be the subset of $\Phi(S,t_r)$ that includes
only followers of $R$.
What is the fraction of these users that 
follow $S$ by time $t_r+\Delta$?

\begin{equation}
	P_{ENDO}(\Delta) = \frac{|L: L\in \Phi_R(S, t_r), L\in F(S, t_r + \Delta)|}{|\Phi_R(S,t_r)|}
\end{equation}

In a small-scale dataset (Dataset-1), we observed 4,945 new followers
for the 200 monitored Speakers over 10 days.
TRF events accounted for 42\% of these new links.
This shows that TRF events are rather infrequent, compared 
to tweets and retweets, but they are responsible
for a large percentage of the new links in Twitter. 

Figure~\ref{fig:ptrfVSptwf} compares the two probabilities for increasing
values of $\Delta$, averaged across all TF and TRF events in our dataset.
We omit confidence intervals because they are too narrow.
Note that the probability of endogenous new followers is
consistently much higher than the probability of exogenous new followers.
Especially for short $\Delta$ (up to 2 hours), $P_{ENDO}$ 
is three orders of magnitude higher than $P_{EXO}$. 
The difference drops to two orders of magnitude and remains stable even 
for values of $\Delta$ larger than 24 hours.

Please note that the previous comparison does not prove causality:
{\em we cannot be certain whether a user $L$ decided to follow
$S$ because she received a retweet of $S$.}
However, if $L$ had not received that retweet it would be 100-1000 times less 
likely that she would follow $S$ within a given time interval.

Figure~\ref{fig:ptrfVSptwf} shows that $P_{ENDO}$ increases significantly
as $\Delta$ increases to about 24 hours. 
After that point, $P_{ENDO}$ saturates to a value that is about $10^{-2}$.
It can be argued that this underestimates the actual TRF probability. 
The reason is that a large fraction of Twitter users are either 
completely inactive or they do not visit Twitter often. 
Recent statistics report that only 20\% of registered users 
visit Twitter at least once per month~\cite{TWstatbrain}.
Additionally, a report from Pew Internet~\cite{PewInternet} in 
2010 reported that only 36\% of Twitter users check their inbox at 
least once a day.

\subsection*{TRF characteristics}
The previous analysis 
verifies our initial intuition that the likelihood with which a user $L$ 
follows a user $S$ greatly increases when $L$ receives a retweet 
of $S$. Furthermore, this likelihood is also affected by the length of 
the interval between the retweet and the time $L$ observed that retweet. 
We now give a more precise definition of Tweet-Retweet-Follow events. 
We say that a Tweet-Retweet-Follow event between 3 users $S$, $R$, and $L$
occurs when we observe the following sequence of events:
\begin{enumerate}[(a)]
\item $S$ tweets a message $M$ at time $t_s$,
\item A user $R$ retweets $M$ at some time $t_r>t_s$,
\item A user $L$, who is a follower of $R$ (i.e. $L \rightarrow R$) 
at $t_r$ but not a follower of $S$, 
follows $S$ by time $t_l$, where $t_l \in [t_r, t_r + \Delta]$.
\end{enumerate}
Note that $R$ may not be a direct follower of $S$. 

We collected a larger dataset (Dataset-2) that we use to analyze 
and model TRF events.
In this dataset we observe 7,451 TRF events,
which represent 17\% of the observed new follower relations. 

\begin{figure}[t]
	\centering
	\subfigure[] {
		\includegraphics[width=0.3\columnwidth,height=0.2\columnwidth]{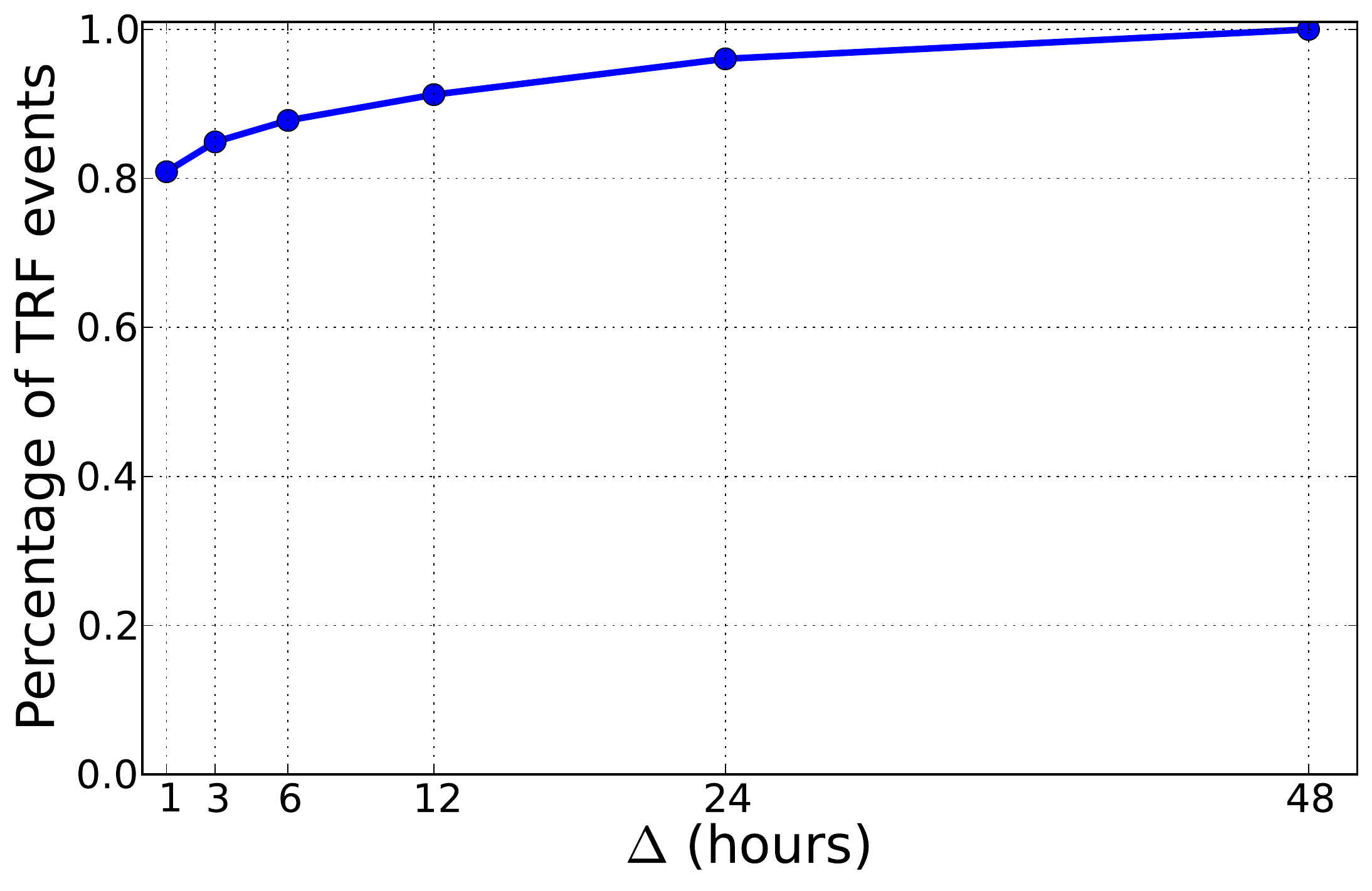}
		\label{fig:trfVsDelta}
	}
	\subfigure[] {
		\includegraphics[width=0.3\columnwidth,height=0.2\columnwidth]{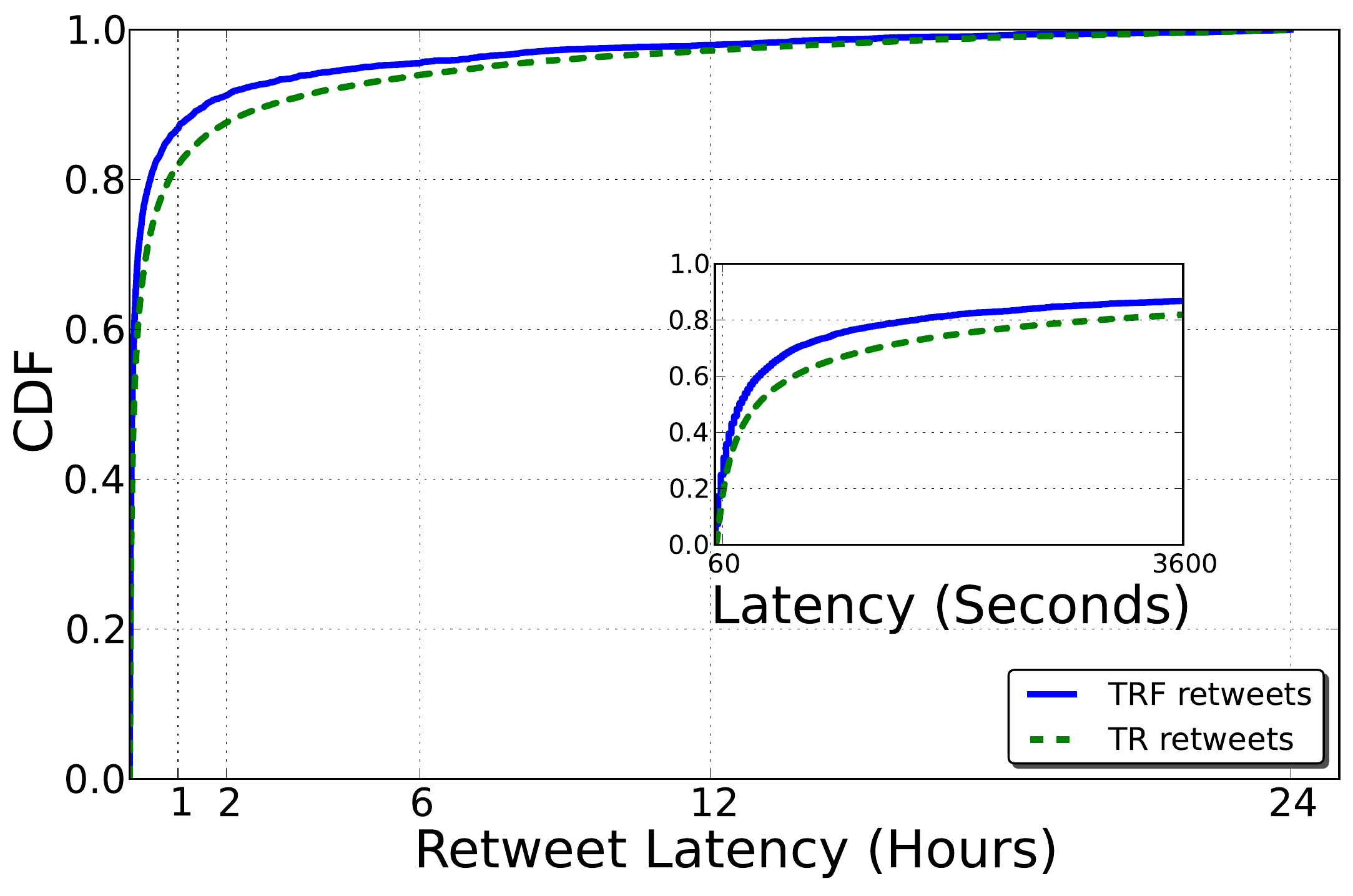}
		\label{fig:rtlatfit}
	}
	\subfigure[] {
		\includegraphics[width=0.3\columnwidth,height=0.2\columnwidth]{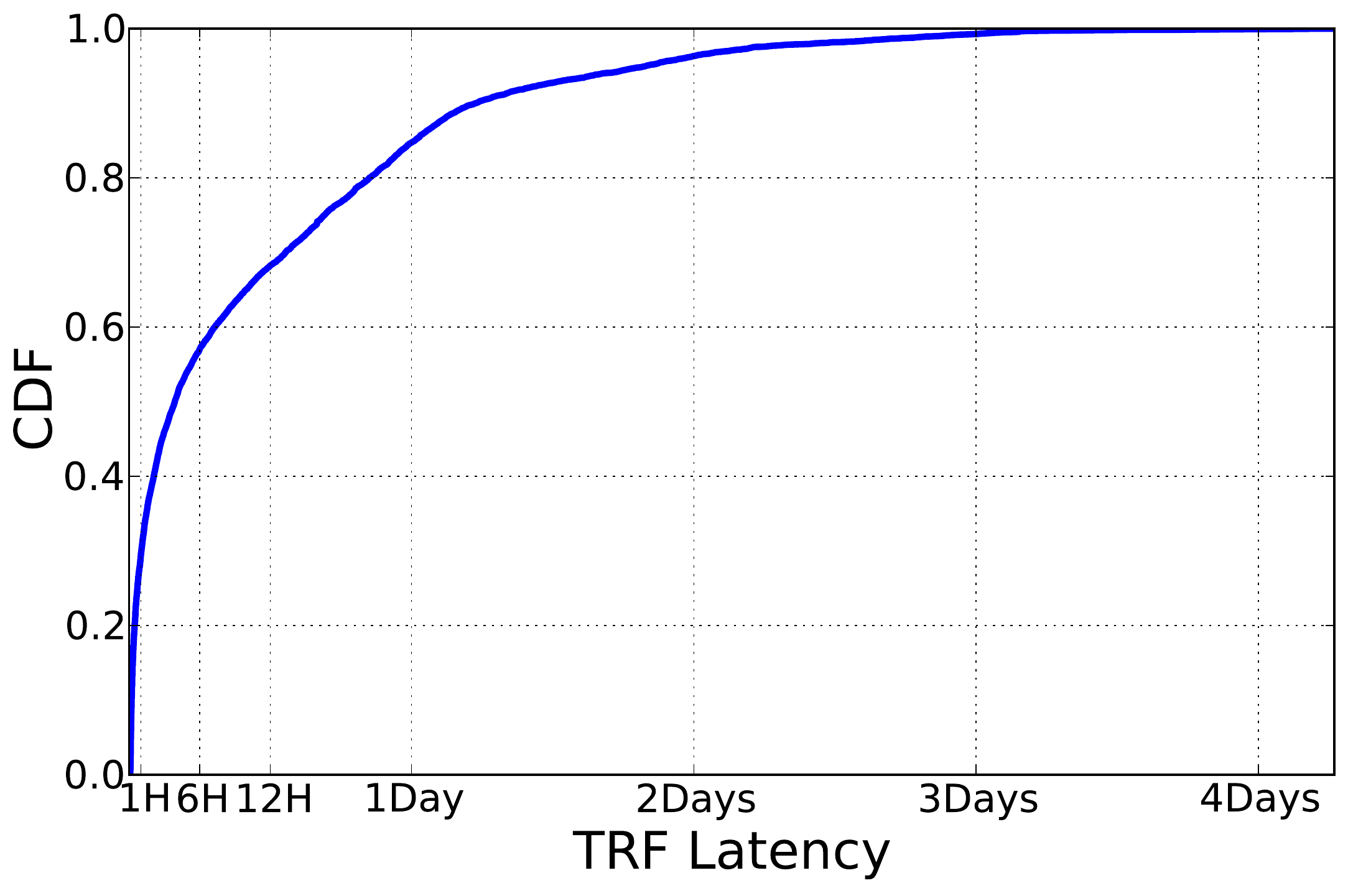}
		\label{fig:tfMtr}
	}
	\caption{~\subref{fig:trfVsDelta} Percentage of identified TRF events as function of $\Delta$. ~\subref{fig:rtlatfit} Retweet latency for all observed retweets. We plot separately retweets that lead to a TRF event, and retweets that do not (TR retweets). ~\subref{fig:tfMtr} Delay between the time of a retweet of Speaker $S$ and the time the Listener $L$ follows $S$.}
	\label{fig:trfAn}
\end{figure}

$\Delta$ is the only parameter in this definition
and it affects the likelihood of TRF events.
Figure~\ref{fig:trfVsDelta} shows the percentage of identified TRF 
events as a function of the parameter $\Delta$.
As expected, the number of TRF events increases with $\Delta$ but
most of them occur within 24 hours from the corresponding retweet.

\textbf{Retweet latency:} 
Figure~\ref{fig:rtlatfit} distinguishes between retweets that 
resulted in at least one TRF event (TRF retweets) 
and retweets that did not result in a TRF event (TR retweets).
The analysis of these retweet events shows that more than 90\% of 
them occur in less than an hour from the corresponding tweet; 
we refer to this time interval as {\em retweet latency}. 
This result supports the idea that ``retweeting users'' tend to act soon 
after new information becomes available.

\textbf{TRF latency:} 
We observe new $L \rightarrow S$ relations even 4 days after 
$L$ has received a retweet of $S$, as shown in Figure~\ref{fig:tfMtr}. 
However, more than 80\% of the TRF events 
occur in less than 24 hours after the retweet. 
Unless stated otherwise, in the rest of this paper we set $\Delta$=24 hours.

\subsection*{TRF probability }
For each monitored Speaker, we collect at each sampling instant 
her list of followers $F(S)$, 
tweets, retweets, Repeaters and the set of followers for each Repeater $F(R)$.
We then identify the set of {\em Tweet-Retweet (TR) events}
for each retweet of Speaker $S$: 
${TR}(S, R, L, t_r, I_{\Delta})$.
A TR event denotes that Listener $L$ received a message of $S$ 
at time $t_r$ through a retweet by Repeater $R$.
The indicator variable $I_{\Delta}$ is  1 if $L$ 
followed $S$ during a time period of length $\Delta$ after $t_r$.

We could define the TRF probability
as the fraction of TR events for which $I_{\Delta}$=1.
This calculation, however, does not consider 
that a Listener may receive multiple retweets 
(of the same or different tweets) of that Speaker.
It would not be realistic to assume that the Listener will 
decide whether to follow the Speaker immediately after each retweet.
Typically, users do not read each tweet immediately
when it is generated, nor they 
have an infinite attention span that would 
allow them to consider all tweets in their inbox~\cite{weng2012competition}.
It is more reasonable to expect that each time a user opens 
her inbox she reads several recent tweets at the same time. 
So, we assume that a Listener decides whether to follow a Speaker
based on a group of received retweets that were recently received.

Specifically, we group TR events into {\em Retweet Groups (RG)} as follows. 
Each RG is represented as ${RG}(S, L, t_r, n, I_{\Delta})$, 
where $S$ and $L$ are the Speaker and Listener, respectively, 
$t_r$ is the timestamp of the first retweet in that group, 
and $n$ is the number of retweets of $S$ received by $L$ 
during the time window $<t_r, t_r + \Delta>$.
Note that these retweets may be generated by different Repeaters.
The indicator variable $I_{\Delta}$ is 1 if $L$ 
followed $S$ by the end of the previous time interval.
If $L$ followed $S$ at time $t_r \le t \le t_r + \Delta$,
the corresponding RG includes only those retweets received by $L$ before $t$;
any subsequent retweets are ignored because $L$ already follows $S$.

Based on this Retweet Grouping method, 
we calculate the TRF probability $P_{TRF}(\Delta)$ as the fraction of 
RGs for which $I_{\Delta}$=1.

\subsection*{Factors that affect the TRF probability}

We now examine a number of factors that may affect the TRF probability.
The small magnitude of the TRF probability 
makes the identification of important factors more challenging~\cite{he2009learning};
the following results, however, are given with satisfactory 
statistical significance (see p-values in Table~\ref{tab:factors}).

Table~\ref{tab:factors} lists the structural and informational 
factors (features) we consider. 
We use logistic regression
to analyze how these features correlate with the TRF probability.
Based on (\ref{eq:lregr}), we estimate the correlation 
coefficient $\kappa_i$ for each factor $X_i$. 
$\kappa_i$ denotes the effect of $X_i$ to the ``odds'' of TRF events,
\begin{equation}\label{eq:lregr}
ln  \left( \frac{P_{TRF}}{1-P_{TRF}} \right) = \kappa_0 + \sum_{i=1}^n \kappa_i\,X_i
\end{equation}
Table~\ref{tab:factors} shows the odds ratio and 
the corresponding 95\% confidence interval for each feature.
An odds ratio $\rho$ represents a 
$\rho \times P_{TRF}$ increase in the TRF probability 
for every unit increase of the corresponding feature.
Thus, odds ratios close to 1 suggest that those features have no 
major effect on the TRF probability. 
Table~\ref{tab:factors} shows that all odds ratios 
are statistically significant ($p < 0.01$).

The ``Twitter age'' of the Speaker, the number of followers 
and followees (factors that were previously shown to 
correlate with Twitter activity) as well as
the tweeting~\cite{huberman2008social,kwak2010twitter} 
and retweeting~\cite{suh2010want} rate of the Speaker, 
show no correlation with the TRF probability. 
Similar results are obtained when examining 
the age and number of followers or followees of the Listener.
We have also examined a number of aggregated informational features, 
namely the Speaker's overall activity and her 
daily tweeting activity.
Both features show no significant correlation with the TRF probability.

\textbf{Reciprocity:} 
A structural feature that examines 
the reverse relation between $S$ and $L$, i.e., {\em whether $S$
was already following $L$ when $L$ received one or more retweets of $S$}, 
has a large effect on the TRF probability.
Reciprocity increases the probability that $L$ will follow $S$ 
by 27.3 times compared to the base TRF probability. 
Previous work has shown {\em reciprocity} to be a dominant 
characteristic of several online social networks such as 
Twitter~\cite{kwak2010twitter}, Flickr~\cite{cha2009measurement}, 
and Yahoo 360~\cite{kumar2010structure}.

\begin{figure}[t]
	\centering
	\includegraphics[width=\columnwidth,height=0.25\columnwidth]{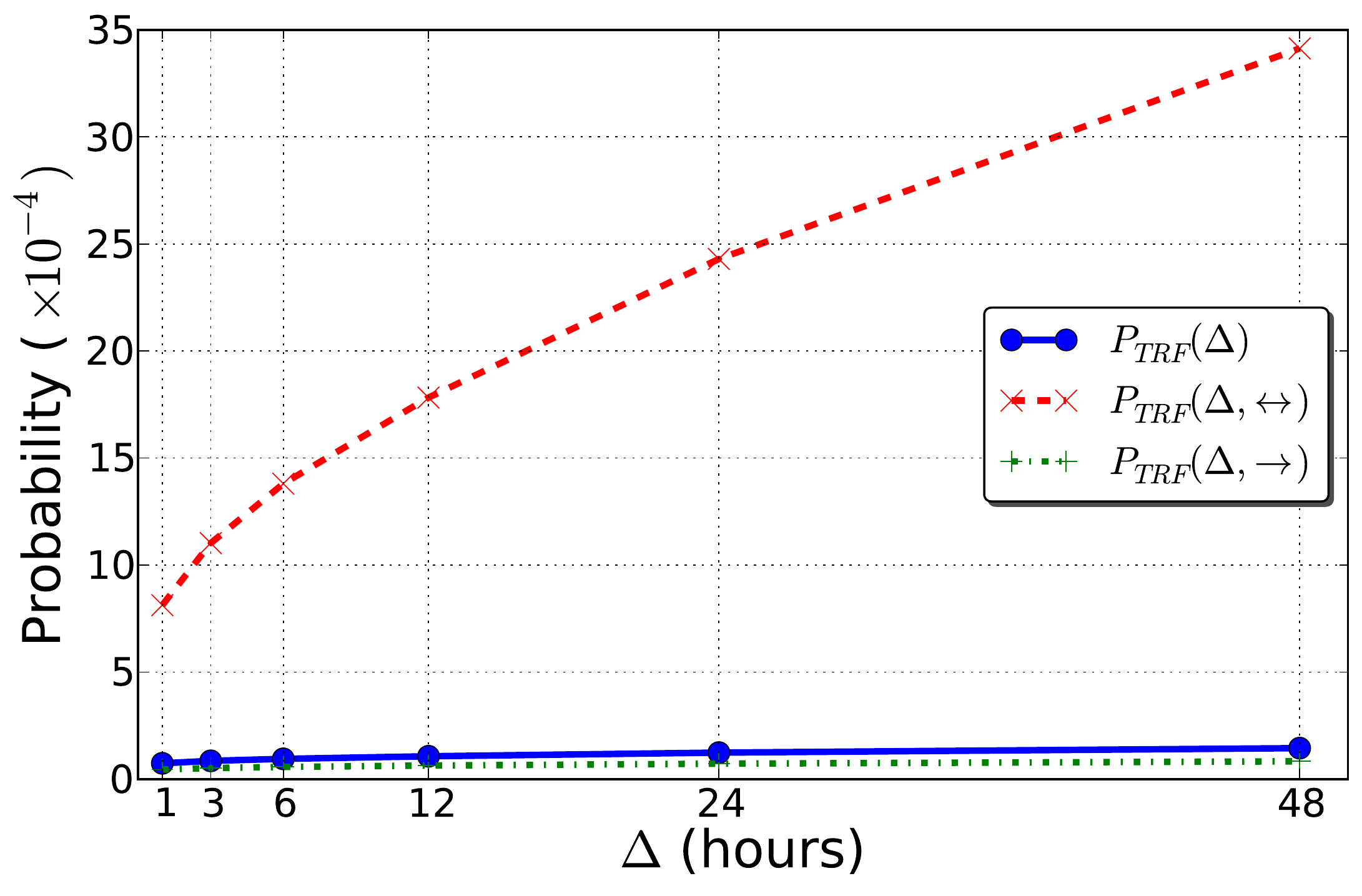}
	\caption{$P_{TRF}(\Delta)$, Reciprocal $P_{TRF}(\Delta,\leftrightarrow)$ and Non-reciprocal $P_{TRF}(\Delta,\rightarrow)$.}
	\label{fig:trfProbVsDelta}
\end{figure}

In 44\% of the observed TRF events, 
the Speaker was following the Listener prior to the formation of the reverse link. 
Figure~\ref{fig:trfProbVsDelta} shows $P_{TRF}(\Delta)$ independent
of reciprocity (solid line), 
when reciprocity is present (dashed line), 
and when reciprocity is not present (dotted line).
When reciprocity is present, the TRF probability,
denoted by $P_{TRF}(\Delta,\leftrightarrow)$, 
is one order of magnitude larger than the probability 
without reciprocity, denoted by $P_{TRF}$ $(\Delta,\rightarrow)$. 
For $\Delta >$ 3 hours, 
$P_{TRF}(\Delta,\leftrightarrow)$ further increases and gradually
becomes up to two orders of magnitude larger.

The large quantitative effect of reciprocity on the TRF probability
implies that there may be different reasons 
for the formation of a link from the Listener to the Speaker
in that case.
The existence of the reverse link, $S \rightarrow L$, could 
imply that these two users have some prior relation.
They may know each other in other social contexts (online or offline) 
or they may belong to similar interest groups. 
In such cases, the retweet of $S$ can make $L$ aware of the existence 
and activity of $S$ in the Twitter network. 

\textbf{Number of tweets and repeaters:} 
Earlier social influence studies showed that
the probability that an individual adopts a new behavior increases 
with the number of her ties already engaging 
in that behavior~\cite{backstrom2006group,bakshy2009social,Hodas:2012:VDA:2411131.2411644,DBLP:conf/www/RomeroMK11}. 
Similarly, we examine whether the number of tweets and retweets of $S$ received by $L$ 
affects the TRF probability. 
It turns out that the TRF probability increases with 
both the number of distinct tweets of $S$ that $L$ receives 
(odds ratio = $2.01$),
and with the number of distinct Repeaters that $L$ received retweets
from (odds ratio = $2.08$). 

For simplicity, 
we choose to aggregate the number of distinct Repeaters 
and the number of distinct tweets of $S$ that $L$ received 
into a single parameter: the total number $n$ of retweets (potentially
not distinct) of $S$ that were received by $L$ in a time period of 
length $\Delta$.
This new factor has high correlation with the TRF probability 
(odds ratio = $1.25$, $p < 0.001$). 
Figure~\ref{fig:h48fit}-top shows the TRF probability 
in the absence of reciprocity ($L \rightarrow S$)
while Figure~\ref{fig:h48fit}-bottom shows the TRF probability
in the presence of reciprocity ($L \leftrightarrow S$),
as a function of $n$.

\subsection*{TRF model}
We now construct a simple model for the probability of TRF events.
The objective of this exercise is to create
a parsimonious probabilistic model that can be used in analytical or 
computational studies of co-evolutionary dynamics in social networks. 

The model considers two independent mechanisms behind each TRF event:
How many retweets $n$ of Speaker $S$ did the Listener $L$ receive?
And second, did $L$ actually observe (i.e., read) this group of retweets? 
The simplest approach is to assume, first, that the $n$ received
retweets are either observed as a group with probability $p$
or they are completely missed, and second, that each observed retweet causes
a TRF event independently and with the same probability $q$. 
Then, the probability of a TRF event after receiving at most $n$ 
retweets  is
\begin{equation}
P_{TRF}(n) = p \times \left(1 - (1-q)^n\right)
\label{eq:nonrpmodel}
\end{equation}
Thus, the probability of a TRF event after only
one received retweet is $p\,q$.
For a large number of received retweets, the TRF probability
tends to the observation probability $p$. 

\begin{figure}[t]
\centering
	\includegraphics[width=\columnwidth,height=0.25\columnwidth]{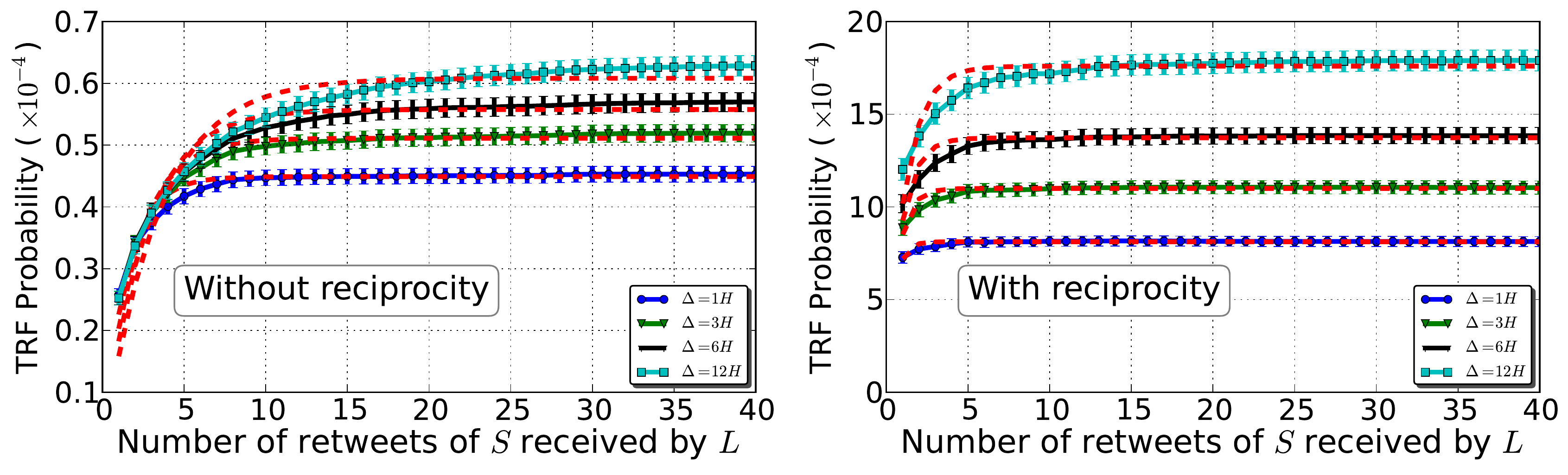}
	\caption{Empirical (solid) and model-based (dashed) TRF probability 
$P_{TRF}(\leftrightarrow, n)$ (left) and 
$P_{TRF}(\rightarrow, n)$ (right) as a function of the number $n$ of received retweets of $S$ at $L$, for four different values of $\Delta$}
	\label{fig:h48fit}
\end{figure}

As shown in Figure~\ref{fig:h48fit}-left,
the measured TRF probability $P_{TRF}(\rightarrow, n)$ without reciprocity
seems to ``saturate'' after $n$ exceeds about 10-20 retweets.   
The same trend is observed in the case of reciprocity
(Figure~\ref{fig:h48fit}-right),
but the saturation appears earlier (after abound 5-10 retweets). 
The model of (\ref{eq:nonrpmodel}) captures the dependency with $n$
quite well. The parameters $p$ and $q$ depend on 
reciprocity as well as on the time window $\Delta$, as shown in
Table~\ref{tab:pandq}.
Reciprocity increases significantly both the observation
probability $p$ and the probability $p \, q$  that a single 
received retweet will cause a TRF event.
As expected, increasing the observation time window $\Delta$ 
increases the observation probability.
The effect of $\Delta$ on the probability $p\,q$ is weaker,
especially when there is no reciprocity.

\section*{Discussion}
\label{sec:trfeffect}

Most prior work in online social networks focused
either on the exogenous evolution of the topology (dynamics of network)
or on influence and information diffusion on static networks (dynamics on network),
ignoring the potential coupling between these two dynamics.
In this paper, we considered co-evolutionary dynamics
in the specific case of the Twitter online social network.
Our study focused on the addition
of new links through the so-called Tweet-Retweet-Follow events.
We showed that TRF events, although infrequent compared to
tweets or retweets, occur in practice and
they are responsible for a significant fraction (about 20\%) of the new edges in Twitter.
Through (near) real-time monitoring of many Twitter users,
we showed how to identify TRF events and
investigated their temporal and statistical characteristics.
More than 80\% of TRF events occur in less than 24
hours after the corresponding retweet.
The main factors that affect the probability of a TRF event
are reciprocity and the total number of retweets received by the Listener.

We now discuss how TRF events may gradually transform the 
structure of a social network. 
We consider two fundamentally different network topologies, 
and discuss the implications of 
TRF events from the information diffusion perspective.

\begin{figure}[t]
\centering
	\subfigure[Circular topology] {
		\includegraphics[width=0.45\columnwidth,height=0.2\columnwidth]{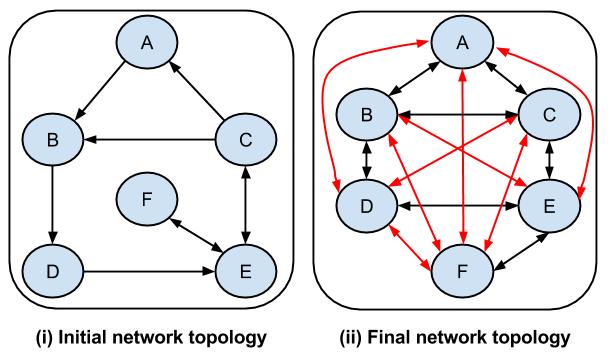}
		\label{fig:circularex}
	}
	\subfigure[Hierarchical topology] {
		\includegraphics[width=0.45\columnwidth,height=0.2\columnwidth]{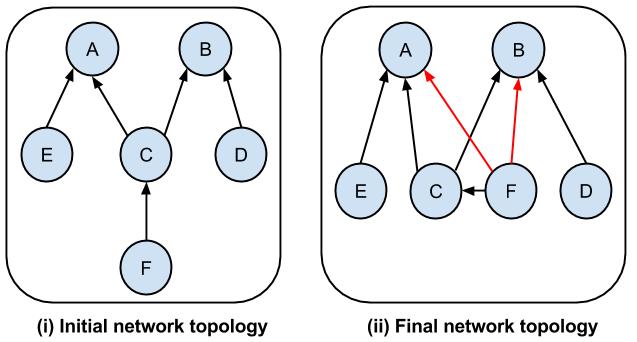}
		\label{fig:hierarchicalex}
	}
	\caption{\subref{fig:circularex} An initial network that includes a directed cycle. 
	A sequence of TRF events can transform this cycle to a clique, meaning
	that the corresponding users gradually form a tightly knit community. 
	~\subref{fig:hierarchicalex} A hierarchical initial network. A sequence of TRF events can
	transform this multi-layer hierarchy into a two-layer hierarchy in which each 
	sink node is directly followed by a set of other nodes 
	(its ``sphere of influence''), while each non-sink node follows at 
	least one sink node.}
	\label{}
\end{figure}

\textbf{Effect on topologies with directed cycles:} 
The left graph of Figure~\ref{fig:circularex} shows a weakly connected network,
which may be a subset of the Twitter topology. 
A directed cycle exists between some of its nodes, namely 
$A \rightarrow B \rightarrow D \rightarrow E \leftrightarrow C \rightarrow A$. 
Let us focus on the largest directed cycle in this network,
i.e., in its largest Strongly Connected Component (SCC).
The ties of the participating nodes may also include links 
to or from nodes out of this cycle, such as the $E \leftrightarrow F$ 
relation in this example. 

Suppose that $A$ posts a tweet at some point in time and that
$C$ decides to retweet it. 
In that case, node $E$ will receive that retweet and may follow $A$
(TRF event). 
It is easy to see that, after a sufficiently large number of TRF events,
the nodes of this directed cycle  
will form a fully connected directed graph, 
as shown in the right graph of Figure~\ref{fig:circularex} 
(red edges denote connections created through TRF events), 
in which everyone is following all others. 
This transformation can only take place when a cycle already exists in 
the initial network; TRF events {\em cannot} create directed cycles.
So, when an initial network includes a directed cycle, a sequence
of TRF events may transform that cycle into a clique 
in which everyone can generate information that all others
receive directly from the source.

\textbf{Effect on hierarchical topologies:} 
The left graph of Figure~\ref{fig:hierarchicalex} 
shows a hierarchical weakly connected directed network.
Again, this network may be a subset of the Twitter topology. 
This network contains no directed cycles, but a number 
of sink nodes (i.e. nodes with no outgoing edges; $A$ and $B$ in this example).

User $F$ may receive a retweet of $A$ and $B$ 
through $C$, and she may then decide to follow them. 
After a sequence of TRF events, this network can then 
reach the topological equilibrium shown in 
the right graph of Figure~\ref{fig:hierarchicalex}, 
in which no new links can be added through TRF events. 
More generally, suppose that $F'(X)=\{X_1,\ldots,X_n\}$ is the set of 
followees of $X$.
The set of Speakers that $X$ may receive a retweet from
can be defined recursively as 
$F'_U(X) = F'(X) \cup (F'_U(X_1) \cup \ldots F'_U(X_n))$;
if user $X$ does not have any followees then $F'_U(X)$ is the empty set.
It is easy to see that, after a sufficiently large number of TRF events, 
a multi-layer hierarchical network will converge to a two-layer hierarchy
in which every non-sink user X follows {\em all} users in $F'_U(X)$. 
Then, an initial sink node $X$ will be followed directly  
by all users that had a directed path towards $X$ in the initial network.
A consequence of TRF events in such hierarchical networks is the 
emergence of some highly influential users that were the sink nodes 
in the initial network.
Further, non-sink nodes will be partitioned, with the
users in each partition following a distinct set of sink nodes.

The previous two topologies are obvious extremes.
In practice, a given weakly connected subset of Twitter users 
may contain groups of nodes that form directed cycles 
as well as nodes that do not belong in any directed cycle. 
An interesting question then is: {\em given a weakly connected 
directed social network, what fraction of its nodes 
belong to the longest directed cycle (i.e., largest SCC) in that network?} 
If this fraction is large, the network resembles the 
example of Figure~\ref{fig:circularex}, 
while if it is close to zero the network is similar to the 
example of Figure~\ref{fig:hierarchicalex}.

We investigated the previous question based on samples
of the actual Twitter topology, at least as it was measured
by Kwak et al.~\cite{kwak2010twitter} in 2010.  
We collected weakly connected network samples using the 
{\em Random-Walk}~\cite{Leskovec:2006:SLG:1150402.1150479} 
and {\em Snowball} (Breadth-First-Search)~\cite{goodman1961snowball} 
sampling methods.
The largest SCC was determined with 
Tarjan's algorithm~\cite{tarjan1972depth}.

In the case of moderately large samples, between 1,000  to 1,000,000 nodes,
{\em the largest SCC contained consistently more than 90\% of the nodes.}
This result suggests that the Twitter topology is closer to the 
network of Figure~\ref{fig:circularex} than to the 
network of Figure~\ref{fig:hierarchicalex}.
The creation of such large cliques, however, may require a very 
long time, and it may also be impractical for a user to follow 
thousands of other users. 
Consequently, we are more interested in smaller samples, 
including only tens or hundreds of Twitter users.

\begin{figure}[t]
	\centering
	\includegraphics[width=\columnwidth,height=0.3\columnwidth]{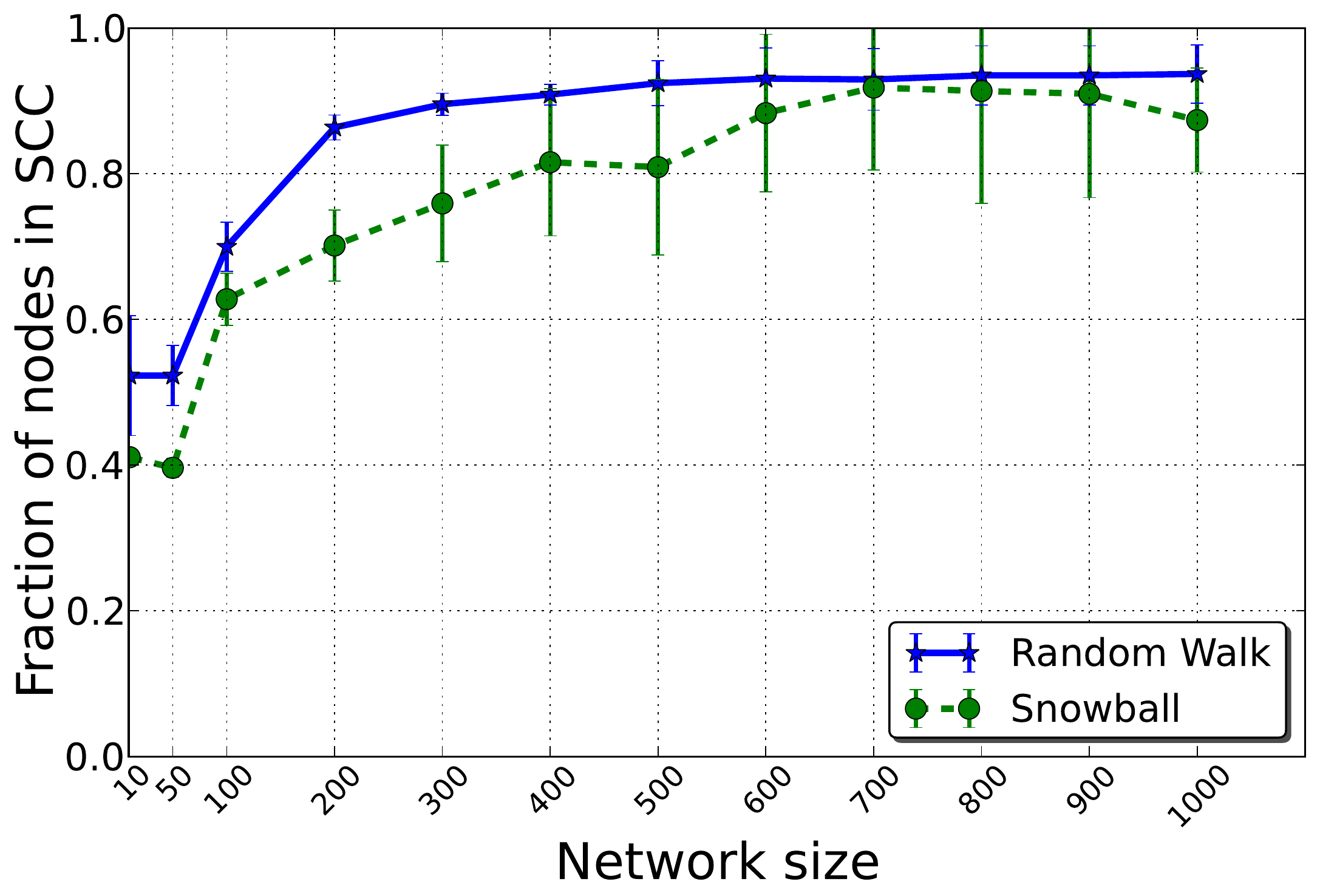}
	\caption{Fraction of Twitter nodes in the largest SCC for different sample sizes, using two sampling methods.}
	\label{fig:sampl}
\end{figure}

Figure~\ref{fig:sampl} shows the percentage of Twitter users
that are included in the SCC of small network samples, 
in the range of 10-1,000 nodes.
Each point is the average of 1,000 samples of that size
and the error bars represent 95\% confidence intervals.
Independent of the sampling method, 
the SCC typically includes the majority of the nodes 
even for samples of few tens of users. 
The SCC percentage increases 
to about 80-90\% for networks with more than 200-400 users. 
These results imply that co-evolutionary dynamics,
and the TRF mechanism in particular, have the 
potential to gradually create very dense communities of users
in which everyone is following almost everyone else,
as long as the involved users are active, tweeting 
and retweeting information.

\section*{Methods}

\subsection*{Data Collection} 
To identify TRF events we need to observe the 
appearance of a new follower link from an arbitrary Listener $L$
to a monitored Speaker $S$, shortly after $L$ has received
a retweet of $S$ through a Repeater $R$.
This requires information about both the time of the retweet(s)
as well as the time the new follower link has appeared.
The Twitter API, though extended in functionality, 
does not provide information about the creation time of follower relations. 
Furthermore, existing link creation time inference methods~\cite{meeder2011we} 
are not applicable in our study because they cannot be used in real-time. 
To retrieve (near) real-time timing we have implemented a 
Twitter data retrieval system that periodically checks 
for new followers and retweets in a given set of Speakers. 
We explain each step of the process in the following paragraphs.

\textbf{Selection of active Speakers:} 
We obtain a number of active Twitter users as potential Speakers 
through a stratified sampling method. 
It has been reported that about 25\% of
Twitter users have never posted any 
messages~\cite{beevole} 
and that most users check their Twitter feeds rarely~\cite{kwak2010twitter}.
A random user selection process would most likely visit a number of users 
without any recent posts, 
wasting a large number of our limited Twitter API calls.
The adopted sampling method ensures that we monitor
users that have recently posted a tweet. 
Specifically, 
we  crawl the Twitter search page~\cite{twsearch} based on 
a single-character search selected at random from the set of $[1-9A-Za-z]$. 
The search page returns 
the latest 20 tweets containing the search term. 
We identify the users 
that posted these tweets and add them to our monitored Speakers set. 
For each selected Speaker, we also collect information about their 
``join time'', number of followees, 
followers and posted tweets. 
For each observed tweet, we collect the time it was posted and 
the posted message.

Given this set of monitored Speakers,
we look for any retweets of their tweets posted during the last two hours.We 
only consider retweets that
are flagged as such by the Twitter API.
For each retweet, we retrieve the set of 
followers and set of followees of the Speaker,
as well as the Repeater $R$ at the time instant we first observed 
that retweet. 
Additionally, we collect the set of followers and followees 
of the Repeater at that time.

\textbf{Monitoring of Speakers:} The previous process results in a 
number of possible TRF events, whenever a 
follower of a Repeater receives a retweet of a monitored Speaker.
To identify new followers we need to examine any changes
in the Speaker's followers before and after the retweet.
To do so, we retrieve the set of followers of the Speaker periodically, 
approximately every 5 minutes.
We identify a TRF event when the set of followers of $S$
gains a new member (the Listener $L$) that 
was previously seen in the set of followers of $R$. 
At that point we log the time that $L$ was seen to follow $S$ 
and calculate the {\em TRF latency} as the time difference 
between the time $R$ retweeted $S$ and the time $L$ followed $S$.
If $L$ received multiple retweets of $S$
(as the same tweet from multiple Repeaters, multiple tweets 
from the same Repeater, or multiple tweets from multiple Repeaters), 
we assign the TRF event to the most recent retweet of $S$ received by $L$. 
The intuition here is that the most recent tweets will appear at the top 
of $L$'s inbox and they are more likely to be read than older retweets.
At this point we also collect the set of followers and followees of the Listener.

Every 5 minutes, we also update the set of monitored Speakers as follows. 
If a selected Speaker has not posted any tweets during the last 24 hours, 
we stop monitoring that user and select a new Speaker using 
the previous sampling method.
The reason is that most new follower relations
tend to occur within few hours from the time a Speaker has been active
~\cite{antoniades2011we,weng2012competition}.

\subsection*{Data collection system} Due to the complexity and the 
real-time nature of our data collection process, we need a large 
Twitter API request throughput. 
We used Twitter's API 1.0, which 
limits users to 350 API requests per hour.
To increase this request throughput we use a large number of 
distributed hosts, provided by PlanetLab, as proxies for accessing Twitter
~\cite{chun2003planetlab}. 
Our collection process is coordinated by a ``dispatcher'' application located at 
Georgia Tech. The dispatcher decides what data are required at any point in 
time and instructs a number of ``workers'' 
to request that data from Twitter. 
Each worker is assigned a single Planetlab host that routes 
API requests to Twitter.
When a worker runs out of requests it deactivates itself 
and notifies the dispatcher. 
At that point the dispatcher generates a new worker, 
providing it with a fresh request workload.

We divide the data collection process to small independent processes, 
each of them requiring the smallest possible number of requests. 
In this way, we partition different parts of the Speaker 
monitoring process to a number of workers, 
speeding up the collection process. 
For instance, when requesting an update for a Speaker, 
the retrieval of tweets, retweets and follower sets are executed 
through different Planetlab hosts.
Further, we limit the number of monitored Speakers, at the same time, to 500
to avoid overloading both Twitter and our collection system.

\subsection*{Bot-filtering} A major concern for any Twitter dataset 
is to avoid bots.
Such accounts act differently than most regular Twitter 
users, biasing the analysis. 
To identify and remove bot accounts from our dataset we revisited
each account three months after the initial data collection to 
check which of those accounts have been suspended by Twitter. 
This practice has been used by Thomas et al.~\cite{thomas2011suspended} 
as ``ground truth'' for the Twitter bot detection problem. 
Further, it has been reported that only few 
bots survive Twitter's policies for more than a week~\cite{sridharan2012twitter}. 
In  our data, about 1\% of the observed users were suspended by Twitter 
(uniformly distributed across Speakers, Repeaters, and Listeners),
accounting for roughly 10\% of the observed TRF events.

\subsection*{Dataset-1} To estimate the exogenous and endogenous probabilities we use a small-scale
dataset (compared to the dataset used in the rest of the paper).
Specifically, 
we monitor 200 unique Twitter users (Speakers) for a period of 10 days. 
For each Speaker, we collect periodically (every 30 minutes) her 
Twitter timeline, tweets and retweets, along with the 
list of her followers.
We also collect the followers of every follower 
of the 200 monitored Speakers. 
Based on this dataset we can observe all Tweet-Retweet (TR) events 
for every monitored Speaker over the course of 10 days, and so we can ask 
whether a Speaker has gained one or more new followers among 
the set of Listeners of her retweets.

\subsection*{Dataset-2} This dataset
was collected during one week, from September 19 to September 25, 2012.
During that period we monitored 4,746 Speakers
that posted 386,980 tweets. 
These messages were retweeted 146,867 times by 83,860 distinct Repeaters. 
After removing bot accounts, we end up with 7,451 observed TRF events.
This figure represents 17\% of the new follower links observed in our dataset.

\begin{table*}[t]
\centering
\begin{tabular}{|l|p{8cm}|l|l|}
\hline
\textbf{Factor} & \textbf{Description} & \textbf{Odds ratio} & \textbf{95\% CI}\\ 
\hline
\multicolumn{4}{|c|}{\textit{Structural Features}}\\
\hline
$|F(S)|$ & Number of followers of  $S$ & $1.000^{***}$ & $[1.000,1.000]$ \\ 
\hline
$|F'(S)|$ & Number of followees of $S$ & $0.999^{***}$ & $[0.999,0.999]$ \\ 
\hline
$AGE(S)$ & Number of days since $S$ joined Twitter & $0.998^{***}$ & $[0.998, 0.998]$ \\ 
\hline
$S \rightarrow L$&  Reciprocity: whether the Speaker was following the Listener at the time  of the TR event & $27.344^{***}$ & $[25.663, 29.136]$ \\ 
\hline
\multicolumn{4}{|c|}{\textit{Informational Features}}\\
\hline
$|ST(S)|$ & Total number of tweets of $S$ & $1.000^{***}$ & $[1.000,1.000]$ \\ 
\hline
$A_{rate}(S)$ & Rate of $S$ tweets per day & $0.989^{***}$ & $[0.988, 0.991]$ \\ 
\hline
$Tweets(S, L, \Delta)$ & Number of distinct tweets of $S$ received by $L$ during period $\Delta$ & $2.010^{***}$ & $[1.781, 2.270]$ \\ 
\hline
$Retweets(S, L, \Delta)$ & Number of distinct retweets of $S$ received by $L$ during period $\Delta$ & $1.603^{***}$ & $[1.371, 1.873]$ \\
\hline
$Repeaters(S, L, \Delta)$ & Number of Repeaters $R$ that $L$ received tweets of $S$ from during period $\Delta$ & $2.076^{***}$ & $[1.889, 2.282]$ \\ 
\hline
\end{tabular}
\caption{List of examined factors.}
\label{tab:factors}
\end{table*}

\begin{table}[t]
	\centering
	\begin{tabular}{|l|l|l|}
	\hline
	\textbf{$\Delta$} & \textbf{$p$} & \textbf{$p \times q$}\\
	\hline
	\multicolumn{3}{|c|}{\textit{Without reciprocity}}\\
	\hline
	1 hour & $0.5 \times 10^{-4}$ &$0.12 \times 10^{-4}$\\
	\hline
	3 hours & $0.5 \times 10^{-4}$ & $0.13 \times 10^{-4}$ \\
	\hline
	6 hours & $0.6 \times 10^{-4}$ & $0.14 \times 10^{-4}$ \\
	\hline
	12 hours & $0.6 \times 10^{-4}$ & $0.15 \times 10^{-4}$ \\ 
	\hline
	24 hours & $0.7 \times 10^{-4}$& $0.16 \times 10^{-4}$\\ 
	\hline
	48 hours & $0.8 \times 10^{-4}$ & $0.16 \times 10^{-4}$\\ 
	\hline
	\multicolumn{3}{|c|}{\textit{With reciprocity}}\\
	\hline
	1 hour & $8.1 \times 10^{-4}$ & $7.2 \times 10^{-4}$\\
	\hline
	3 hours & $11.0 \times 10^{-4}$ & $8.5 \times 10^{-4}$\\
	\hline
	6 hours & $13.0 \times 10^{-4}$ & $9.3 \times 10^{-4}$\\
	\hline
	12 hours & $17.6 \times 10^{-4}$ & $9.3 \times 10^{-4}$\\
	\hline
	24 hours & $24.0 \times 10^{-4}$ & $10.2 \times 10^{-4}$\\
	\hline
	48 hours & $33.1 \times 10^{-4}$ & $10.2 \times 10^{-4}$ \\
	\hline
	\end{tabular}
	\caption{Estimated value of the two model parameters $p$ and $p \times q$}
	\label{tab:pandq}
\end{table}


\begin{thebibliography}{10}
\bibitem{bakshy2009social}
{Bakshy, E., Karrer, B., Adamic, L.A.}
\newblock {Social influence and the diffusion of user-created content}.
\newblock Proceedings of the {\it 10th ACM conference on Electronic
  commerce}, 325--334, (2009).

\bibitem{vespignani2011modelling}
{Vespignani, A.}
\newblock {Modelling dynamical processes in complex socio-technical systems}.
\newblock {\it Nat. Phy.}, {\bf 8}(1):32--39, (2011).

\bibitem{Leskovec:2005:GOT:1081870.1081893}
{Leskovec, J., Kleinberg, J., Faloutsos, C.}
\newblock {Graphs over time: densification laws, shrinking diameters and
  possible explanations}.
\newblock Proceedings of the {\it 11th ACM SIGKDD Int. Conf. on Knowledge
  discovery in data mining}, 177--187, (2005).

\bibitem{gross2008adaptive}
{Gross, T., Blasius, B.}
\newblock {Adaptive coevolutionary networks: a review}.
\newblock {\it J. R. Soc. Interface}, {\bf 5}(20):259--271, (2008).

\bibitem{DBLP:conf/sdm/LeskovecMFGH07}
{Leskovec, J., McGlohon, M., Faloutsos, C., Glance, N.S., Hurst, M.}
\newblock {Patterns of cascading behavior in large blog graphs}.
\newblock {Proceedings of the {\it 7th SIAM Int. Conf. on Data Mining,
  Minneapolis, Minnesota, USA}}, (2007).

\bibitem{kozma2008consensus}
{Kozma, B., Barrat, A.}
\newblock {Consensus formation on adaptive networks}.
\newblock {\it Phys. Rev. E}, {\bf 77}(1):016102, (2008).

\bibitem{PhysRevE.81.046120}
{Shaw, L.B., Schwartz, I.B.}
\newblock {Enhanced vaccine control of epidemics in adaptive networks}.
\newblock {\it Phys. Rev. E}, {\bf 81}:046120, (2010).

\bibitem{volz2009epidemic}
{Volz, E., Meyers, L.A.}
\newblock {Epidemic thresholds in dynamic contact networks}.
\newblock {\it J. R. Soc. Interface}, {\bf 6}(32):233--241, (2009).

\bibitem{rocha2011simulated}
{Rocha, L.E.C., Liljeros, F., Holme, P.}
\newblock {Simulated epidemics in an empirical spatiotemporal network of 50,185
  sexual contacts}.
\newblock {\it PLoS comp. bio.}, {\bf 7}(3):e1001109, (2011).

\bibitem{PhysRevLett.109.238701}
{Perra, N., Baronchelli, A., Mocanu, D., Gon\ifmmode \mbox{\c{c}}\else
  \c{c}\fi{}alves, B., Pastor-Satorras, R., Vespignani, A.}
\newblock {Random walks and search in time-varying networks}.
\newblock {\it Phys. Rev. Lett.}, {\bf 109}:238701, (2012).

\bibitem{Weng:2013:RID:2487575.2487607}
{Weng, L. et al.}
\newblock {The role of information diffusion in the evolution of social
  networks}.
\newblock Proceedings of the {\it 19th ACM SIGKDD Int. Conf. on Knowledge
  discovery and data mining}, 356--364, (2013).

\bibitem{guptawtf}
{Gupta, P., Goel, A., Lin, J., Sharma, A., Wang, D., Zadeh, R.}
\newblock {WTF: The who to follow service at Twitter}.
\newblock Proceedings of the {\it 22nd Int. Conf. on World Wide Web}, 
  505--514, (2013).

\bibitem{TWstatbrain}
statisticbrain.com.
\newblock {Twitter Statistics}.
\newblock \url{http://www.statisticbrain.com/twitter-statistics/}, (2013).
\newblock [Online; accessed 30-Jan-2014].

\bibitem{PewInternet}
{Smith, A., Rainie, L.}
\newblock {8\% of online Americans use Twitter}.
\newblock
  \url{http://www.pewinternet.org/Reports/2010/Twitter-Update-2010.aspx}.
\newblock [Online; accessed 30-Jan-2014].

\bibitem{weng2012competition}
{Weng, L., Flammini, A., Vespignani, A., Menczer, F.}
\newblock {Competition among memes in a world with limited attention}.
\newblock {\it Sci. Rep.}, {\bf 2}, (2012).

\bibitem{he2009learning}
{He, H., Garcia, E.A.}
\newblock {Learning from imbalanced data}.
\newblock {\it IEEE T. Know. Data En.}, {\bf 21}(9):1263--1284, (2009).

\bibitem{huberman2008social}
{Huberman, B., Romero, D., Wu, F.}
\newblock {Social networks that matter: Twitter under the microscope}.
\newblock {\it SSRN 1313405}, (2008).

\bibitem{kwak2010twitter}
{Kwak, H., Lee, C., Park, H., Moon, S.}
\newblock {What is Twitter, a social network or a news media?}
\newblock Proceedings of the {\it 19th Int. Conf. on World Wide Web}, 
  591--600, (2010).

\bibitem{suh2010want}
{Suh, B., Hong, L., Pirolli, P., Chi, E.H.}
\newblock {Want to be retweeted? large scale analytics on factors impacting
  retweet in Twitter network}.
\newblock Proceedings of the {\it 2nd IEEE Int. Conf. on Social Computing},
  177--184, (2010).

\bibitem{cha2009measurement}
{Cha, M., Mislove, A., Gummadi, K.P.}
\newblock {A measurement-driven analysis of information propagation in the
  Flickr social network}.
\newblock Proceedings of the {\it 18th Int. Conf. on World Wide Web}, 
  721--730, (2009).

\bibitem{kumar2010structure}
{Kumar, R., Novak, J., Tomkins, A.}
\newblock {Structure and evolution of online social networks}.
\newblock {\it Link Mining: Models, Algorithms, and Applications}, 
  337--357, (2010).

\bibitem{backstrom2006group}
{Backstrom, L., Huttenlocher, D., Kleinberg, J., Lan, X.}
\newblock {Group formation in large social networks: membership, growth, and
  evolution}.
\newblock Proceedings of the {\it 12th ACM SIGKDD Int. Conf. on Knowledge
  discovery and data mining}, 44--54, (2006).

\bibitem{Hodas:2012:VDA:2411131.2411644}
{Hodas, N.O., Lerman, K.}
\newblock {How visibility and divided attention constrain social contagion}.
\newblock Proceedings of the {\it ASE/IEEE Int. Conf. on Social Computing
  and ASE/IEEE Int. Conf. on Privacy, Security, Risk and Trust}. IEEE Computer
  Society, (2012).

\bibitem{DBLP:conf/www/RomeroMK11}
{Romero, D.M., Meeder, B. Kleinberg, J.M.}
\newblock {Differences in the mechanics of information diffusion across topics:
  idioms, political hashtags, and complex contagion on Twitter}.
\newblock Proceedings of the {\it 20th Int. Conf. on World Wide Web}, 
  695--704, (2011).

\bibitem{Leskovec:2006:SLG:1150402.1150479}
{Leskovec, J., Faloutsos, C.}
\newblock {Sampling from large graphs}.
\newblock Proceedings of the {\it 12th ACM SIGKDD Int. Conf. on Knowledge
  discovery and data mining}, 631--636, (2006).

\bibitem{goodman1961snowball}
{Goodman, L.A.}
\newblock {Snowball sampling}.
\newblock {\it An. Math. Stat.}, {\bf 32}(1):148--170, (1961).

\bibitem{tarjan1972depth}
{Tarjan, R.}
\newblock {Depth-first search and linear graph algorithms}.
\newblock {\it SIAM J. Comp.}, {\bf 1}(2):146--160, (1972).

\bibitem{meeder2011we}
{Meeder, B., Karrer, B., Sayedi, A., Ravi, R., Borgs, C., Chayes, J.}
\newblock {We know who you followed last summer: inferring social link creation
  times in Twitter}.
\newblock Proceedings of the {\it 20th Int. Conf. on World wide web}, 
  517--526, (2011).

\bibitem{beevole}
{An Exhaustive Study of Twitter Users Across the World}.
\newblock \url{http://www.beevolve.com/twitter-statistics/}.
\newblock [Online; accessed 30-Jan-2014].

\bibitem{twsearch}
{Twitter search page}.
\newblock \url{http://search.twitter.com}.
\newblock [Online; accessed 30-Jan-2014].

\bibitem{antoniades2011we}
{Antoniades, D. et al.}
\newblock {we. b: The web of short URLs}.
\newblock Proceedings of the {\it 20th Int. Conf. on World wide web}, 
  715--724, (2011).

\bibitem{chun2003planetlab}
{Chun, B. et al.}
\newblock {Planetlab: an overlay testbed for broad-coverage services}.
\newblock {\it ACM SIGCOMM Comp. Comm. Rev.}, {\bf 33}(3):3--12, (2003).

\bibitem{thomas2011suspended}
{Thomas, K., Grier, C., Song, D., Paxson, V.}
\newblock {Suspended accounts in retrospect: An analysis of Twitter spam}.
\newblock Proceedings of the {\it ACM SIGCOMM Internet measurement
  conference}, 243--258, (2011).

\bibitem{sridharan2012twitter}
{Sridharan, V., Shankar, V., Gupta, M.}
\newblock {Twitter games: how successful spammers pick targets}.
\newblock Proceedings of the {\it 28th Annual Computer Security Applications
  Conference}, 389--398, (2012).


\end{thebibliography}
\end{document}